\documentclass[wcp]{jmlr}

\usepackage{longtable}% for long tables
\usepackage{graphicx}

\usepackage{algorithm}
\usepackage{algorithmic}
\usepackage[T1]{fontenc}
\usepackage[utf8]{inputenc}
\usepackage{babel}
\usepackage[font=small,labelfont=bf]{caption}
\usepackage{csvsimple}
\usepackage{breakcites}
\usepackage{booktabs}
\makeatletter
\let\Ginclude@graphics\@org@Ginclude@graphics 
\makeatother

\jmlrvolume{}
\jmlryear{}
\jmlrworkshop{}

\title[AMAT to Improve Adversarial Robustness of DNNs]{Adaptive Adversarial Training to Improve Adversarial Robustness of DNNs for Medical Image Segmentation and Detection}

  \author{\Name{Linhai Ma} \Email{l.ma@miami.edu}\\
  \Name{Liang Liang} \Email{liang@cs.miami.edu}\\
  \addr Department of Computer Science\\
  \addr University of Miami, Coral Gables FL 33146, USA
 }

\begin{document}

\maketitle

\begin{abstract}

It is known that Deep Neural networks (DNNs) are vulnerable to adversarial attacks, and the adversarial robustness of DNNs could be improved by adding adversarial noises to training data (e.g., the standard adversarial training (SAT)). However, inappropriate noises added to training data may reduce a model's performance, which is termed the trade-off between accuracy and robustness. This problem has been sufficiently studied for the classification of whole images but has rarely been explored for image analysis tasks in the medical application domain, including image segmentation, landmark detection, and object detection tasks. In this study, we show that, for those medical image analysis tasks, the SAT method has a severe issue that limits its practical use: it generates a fixed and unified level of noise for all training samples for robust DNN training. A high noise level may lead to a large reduction in model performance and a low noise level may not be effective in improving robustness. To resolve this issue, we design an adaptive-margin adversarial training (AMAT) method that generates sample-wise adaptive adversarial noises for robust DNN training. In contrast to the existing, classification-oriented adversarial training methods, our AMAT method uses a loss-defined-margin strategy so that it can be applied to different tasks as long as the loss functions are well-defined. We successfully apply our AMAT method to state-of-the-art DNNs, using five publicly available datasets. The experimental results demonstrate that: (1) our AMAT method can be applied to the three seemingly different tasks in the medical image application domain; (2) AMAT outperforms the SAT method in adversarial robustness; (3) AMAT has a minimal reduction in prediction accuracy on clean data, compared with the SAT method; and (4) AMAT has almost the same training time cost as SAT. Please contact the authors for the source code.
\end{abstract}

\section{Introduction}

\begin{center}
\begin{minipage}[ht]{\textwidth}
\includegraphics[width=0.98\linewidth]{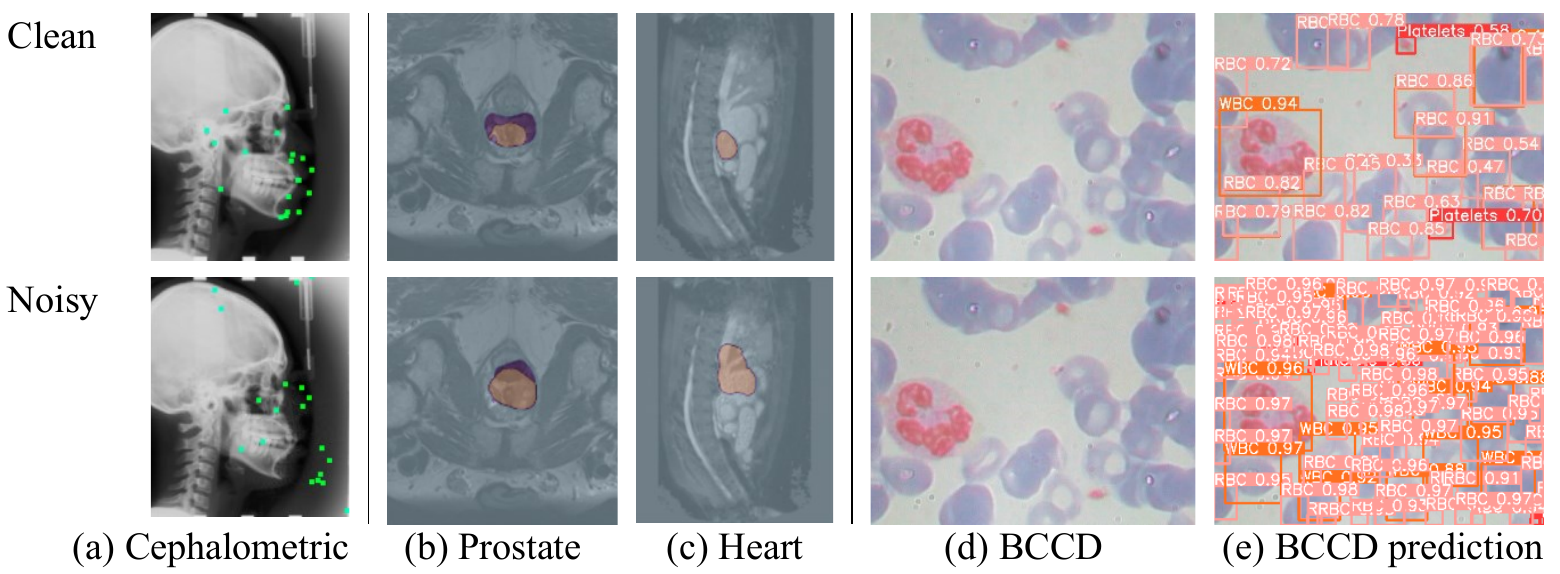}
    \captionof{figure}{Examples of adversarial attacks against cephalometric landmark detection (a), segmentation of prostate (b) and heart (c), and blood cell detection ((d) is the input image and (e) is the detection result). By adding adversarial noises to the original/clean images (the first row), the noisy images are obatained (the second row). DNN outputs are significantly changed when the input images contain adversarial noises.}
    \label{fig1}
 \end{minipage}
 \end{center}

Recent methods based on Deep Neural Networks (DNNs) have achieved high accuracy for medical image analysis, including three basic tasks: segmentation, landmark detection, and object detection. However, DNNs are vulnerable to imperceptible adversarial noises \cite{szegedy2014intriguing, carlini2017towards, brendel2017decision, croce2020reliable, joel2021adversarial, moosavi2016deepfool, ozbulak2019impact, madry2018towards, paschali2018generalizability} (see Fig. \ref{fig1}), which makes it dangerous to deploy DNN-based decision-making systems for life-critical applications, such as medical image-based diagnosis. To make DNNs robust to adversarial noises, adding adversarial noises to training data (i.e., adversarial training) is an effective and general strategy \cite{goodfellow2014explaining,madry2018towards}. However, inappropriate noises added to training data may reduce a model's performance, which is termed the trade-off between accuracy and robustness. For the classification of whole images (e.g., the MNIST dataset), this problem has been sufficiently studied and a tremendous number of defense methods have been proposed to resolve this problem: TRADES \cite{zhang2019theoretically} and MART \cite{wang2019improving} optimize a loss with a KL divergence-based regularization term to make a trade-off between adversarial robustness and standard accuracy. DAT \cite{wang2019convergence} uses convergence quality as a criterion to adjust adversarial training perturbations. ATES\cite{sitawarin2020improving} and CAT \cite{cai2018curriculum} apply curriculum strategy for adversarial training. IAAT\cite{balaji2019instance}, FAT\cite{zhang2020attacks}, Customized AT\cite{cheng2020cat}, MMA\cite{ding2019mma} and IMA\cite{ma2020increasing} are adaptive adversarial training methods and adjust the noise added to training data in the training process based on the classification-aware strategy. GAIRAT \cite{zhang2021geometry} applies sample-wise weights to the loss for robust training and the weights are related to the decision boundary. 

As a comparison, for image segmentation, landmark detection, and object detection tasks in the medical image application domain, the trade-off between accuracy and robustness has rarely been explored and the number of defense methods \cite{liu2020defending, he2019non, daza2021towards} pales. Some of these defense methods are either not effective enough \cite{uesato2018adversarial, liu2020defending}, or only work for very limited network structures \cite{he2019non}.

To improve adversarial robustness of DNNs for classification-related tasks, the standard adversarial training (SAT) method \cite{goodfellow2014explaining, kurakin2016adversarial, madry2018towards} is often considered \cite{daza2021towards}: given a fixed noise level $\varepsilon$ (Lp vector norm), generate a noisy/adversarial sample ($\tilde{x}$, $y$) using the PGD algorithm in \cite{madry2018towards}, where $\tilde{x}=PGD(x,y,\varepsilon)$ and $||x-\tilde{x}||_p\leq\varepsilon$; for every sample ($x$, $y$) in the training set; use the clean data \{($x$, $y$)\} and adversarial/noisy data \{($\tilde{x}$, $y$)\} together to train the model. The SAT loss function is given in \cite{goodfellow2014explaining}:
\begin{equation}
L_{SAT}=(L(f(x),y)+L(f(\tilde{x}),y))/2
\end{equation}
where $L$ is the original loss function for training the DNN model $f$. However, as will be shown in this study, the SAT method is problematic.

In our work to improve the adversarial robustness of DNNs for the three basic tasks (image segmentation, landmark detection, and object detection), we have made the following contributions. 
(1) We investigate the problem of SAT in the medical image segmentation, landmark detection, and object detection tasks: when the training noise level $\varepsilon$ increases, the performance of SAT-trained models improves on data with larger noises, but drops on clean data or data with smaller noises, which indicates that a fixed and unified level of noise for training is inadequate. 
(2) To solve this problem, we design and implement an adaptive-margin adversarial training (AMAT) method for medical image segmentation, landmark detection, and object detection. AMAT generates sample-wise adaptive noises for robust DNN training, instead of using a fixed and unified noise level. In this way, the trained DNN model can have good resistance to adversarial noises while maintaining good accuracy on clean data. Furthermore, AMAT uses a loss-defined-margin strategy so that it can be applied to three different tasks of image segmentation, landmark detection, and object detection.
(3) We successfully apply our AMAT method to improving adversarial robustness of three state-of-the-art DNNs for medical image segmentation, landmark detection, and object detection tasks. Five publicly available medical datasets are used for evaluation. The experimental results demonstrate that: (a) our AMAT method can be applied to the three fundamental tasks in the medical image application domain; (b) AMAT outperforms the SAT method in adversarial robustness on noisy data; (c) AMAT has a minimal reduction in prediction accuracy on clean data, compared with SAT; and (d) AMAT has almost the same training time cost as SAT.

\section{Methodology}

\subsection {Adaptive Margin Adversarial Training (AMAT)}

In our AMAT method, the concept of "margin" is borrowed from support-vector machine \cite{cortes1995support}, which means the distance between a data sample and the decision boundary of a classifier. To improve classifier robustness, during training, the generated noisy/adversarial sample should not cross the true decision boundary (i.e., $\tilde{x}$ and $x$ have the same true label $y$). Since the clean samples have different distances to the true decision boundary, the noisy samples should contain different levels of noises. In this study, although the three tasks are not simple classification of whole images, the margin concept can be generalized: the "margin" of a sample $x$ is defined as the maximum level of noise that can be added to the sample without harming the DNN performance on clean data. The basic idea of AMAT is to let each training sample have its own "margin", instead of using a fixed noise level as SAT.

\begin{algorithm}[t]
%\scriptsize
    \caption{Robust Training in One Epoch}
    \label{algo1}
    \textbf {Input:} \\
    $S$ is the training set.\\
    $f$ is the DNN model. \\
    $Loss$ is the loss function for training the model $f$.\\
    \textbf {Parameters:} \\
    $\mathcal{E}$ is the array of the estimated sample margins. $\mathcal{E}(i)$ is the margin of the sample indexed by the unique ID $i$. Every $\mathcal{E}(i)$ is initialized to be $\Delta_\varepsilon$. \\
    $\Delta_\varepsilon$ is the expansion step size (positive scalar). \\
    $\xi$ is threshold and it is determined by using the loss values on clean data (see Section 2.2).\\
  \textbf{Output:} Updated model $f$ after this training epoch.\\
  \textbf{Process:} 
    \begin{algorithmic}[1]
    \FOR {each training sample ($x$, $y$) with ID $i$ in $S$}   
    \STATE Run the model $f$ on clean sample: $\hat{y} = f(x)$ 
    \STATE Generate a noisy sample using the PGD algorithm: $\tilde{x} = PGD(x, y,\mathcal{E}(i))$ 
    \STATE Run the model $f$ on noisy sample: ${\tilde{y}}=f(\tilde{x})$\\
    \STATE  $L_0=Loss(\hat{y},\ y)$   
    \STATE  $L_1=Loss({\tilde{y}},\ y)$   
    \STATE  $L=(L_0+ L_1)/2 $
    \STATE  Back-propagate from $L$ and update the model $f$  
	\IF{$L_1$ < $\xi$}		
		\STATE $\mathcal{E}(i)=\mathcal{E}(i)+\Delta_\varepsilon$  (Enlarge the margin)
	\ELSE
		\STATE $\mathcal{E}(i)=(||x-\tilde{x}||_p + \mathcal{E}(i))/2$  (Refine the margin)
	\ENDIF
    \ENDFOR \\
    	
	\textbf{Note:} 
	The algorithm runs in mini-batches. $||.||_p$ denotes vector Lp norm.
    \end{algorithmic} 	
\end{algorithm}
The AMAT training process is shown in Algorithm \ref{algo1}, which includes two sub-processes: (1) compute the loss and update the DNN model (Line 2 to 8); and (2) update the  sample margins (Line 9 to 12). $x$ is the input (clean) image, $\hat{y}$ is the DNN output (e.g., segmentation map), and $i$ is the unique ID for the clean sample ($x$, $y$). Given the sample $i$ and its estimated margin $\mathcal{E}(i)$, PGD \cite{madry2018towards} is used to generate a noisy sample $\tilde{x}$ (Line 3). The model $f$ runs on the clean and noisy samples to generate the outputs. Then, the loss $L_0$ on clean data and the loss $L_1$ on noisy data are calculated and combined. Finally, back-propagation runs to update the model. It is important not to add too much noise, because large noise may harm the DNN performance on clean data. To this end, we add the condition $L_1 < \xi$ to control the noise. Generally, the loss $L_1$ on the noisy sample will increase if the noise level increases. In order to preserve DNN performance on clean data, the loss $L_1$ should not be significantly larger than the average loss value on clean data. We note that our novel strategy enables loss-defined-margin for each individual sample, and it also enables our AMAT method to handle different tasks as long as the loss functions are well-defined.

\subsection{Value of $\xi$} 
\label{2.2}

The threshold $\xi$ is used to determine if the noise is too large to be added on a sample for training. Assuming that on the clean training set $S = \{(x_i,y_i)|i=1,2,...\}$, the DNN model $f$ has been well trained; and $Loss$ is the loss function for training $f$ on $S$. Then, we can obtain the loss values of all training samples, denoted by $V = \{Loss(f(x),y)| (x, y) \in S\}$.  In our experiments, we observe that the distribution of the loss values approximately follows Normal distribution. The mean loss value $mean(V)$ and the standard deviation of the loss values $std(V)$ can be obtained. Then, the threshold $\xi$ can be $mean(V)$ or $mean(V)+2\cdot std(V)$, and the choice is dependent on the dataset.

\subsection{Loss function with multiple items} 

For some DNN models, the training loss function contains multiple loss terms. For instance, YOLO v5 \cite{glennjocher20226222936} (used in Section 3.3) has three loss terms. If the loss function contains $K$ loss terms, a small modification is made to Algorithm \ref{algo1}: (1) The threshold $\xi$ becomes a vector of $K$ scalars, and each scalar is calculated by using each of the $K$ distributions of the loss terms on clean data; (2) Line 9 is modified to:
\begin{equation}
\textbf{if} {(L_1[1] < \xi[1]) \& (L_1[2] < \xi[2]) \& ... \& (L_1[K] < \xi[K])} \textbf{then}
\end{equation}

In short, all of the loss terms are considered separately. $mean$ and $std$ (see Section \ref{2.2}) are computed for each loss term. When all of the loss terms on noisy data are smaller than their corresponding thresholds, the noise is acceptable for training.

\section{Experiment Configuration}
\label{3}

\subsection{Basic settings}
All of the experiments are conducted on a server with Nvidia Tesla V100 GPU and Intel(R) Xeon(R) E5-2698 v4 CPU processor (2.20GHz). 
Three state-of-the-art DNN models and five publicly available medical image datasets are used for method evaluation: three for segmentation, one for landmark detection, and one for object detection.

\subsection{Evaluated defense methods}
In the experiments, the model trained only on clean data by the regular standard training method is named "STD"; the model trained by the SAT method with the noise level of $\varepsilon$ is named "SAT$\varepsilon$"; the model trained by our AMAT method is called "AMAT". To our knowledge, there are no other adaptive adversarial training methods for image segmentation, landmark detection and object detection in medical image application domain. The SAT method is still most widely used in many applications. Except for SAT, the existing well-known adversarial training methods are designed for the classification of a whole image (e.g., MNIST digit images with the size of 28x28), and these methods cannot be applied to medical image segmentation, landmark detection, or object detection tasks in this study. Thus, we only use SAT with different $\varepsilon$ as the baseline method in the experiments.

\subsection{Method Evaluation}
PGD-based adversarial attack is widely used for adversarial defense method evaluation \cite{uesato2018adversarial, tramer2020adaptive}. The noises for evaluation are generated by PGD whitebox attack \cite{madry2018towards} with 100 iterations (100-PGD). We also applied IFGSM whitebox attack \cite{kurakin2018adversarial} with 10 iterations (10-IFGSM).The maximum attack noise level in the evaluation is selected such that the prediction accuracy drops to almost 0 (except for Landmark Detection).

\subsection{L2 norm is used to measure noise level}
It has been observed that adversarial noises have inner structures and patterns (see Fig. 1 in \cite{goodfellow2014explaining}, Fig. 1 in \cite{yao2020miss}, Fig. 2 in \cite{akhtar2018threat}), which is completely ignored by L-inf norm. This means L-inf is not a good measure of adversarial noises. Thus, in this paper, the adversarial noises are measured in the L2 norm.

\subsection{Evaluation on image segmentation}

In this experiment, we apply SAT and AMAT to a self-configuring DNN, nnUnet \cite{isensee2021nnu}. The nnUnet can automatically configure itself, including preprocessing, network architecture, training, and post-processing for the dataset. The inputs of nnUnet are 2D slices of 3D images. Some of the image slices have poor quality, and the nnUnet does not filter them out, which leads to a big $std(V)$ (see Section 2.2) in the training set. As a result, in this experiment, for AMAT, $\xi$ is set to $mean(V)$ (see Section 2.2). The "Average Dice Index (ADI)" is used as the metric, whose formula is:
\begin{equation}
ADI = \frac{1}{n} \times  \sum_{i=1}^n{\frac{2\times TP_i}{2\times TP_i + FP_i +FN_i}} 
\end{equation}
Here, $n$ is the number of samples in the test set. For the sample $i$, $TP_i$ is the number of pixels in true-positive area, $FP_i$ is the number of pixels in false-positive area, and $FN_i$ is the number of pixels in false-negative area. The metric used in the evaluation of the nnUnet \cite{isensee2021nnu} is renamed "Total voxel Dice Index (TVDI)" in this paper, and it does not equally weight the samples. TVDI is always higher than ADI. We note that TVDI results are in the Appendix, and some are in Table \ref{tdvi}. We use this TVDI metric to show that the clean TVDI values of our results are similar to the results reported in the original nnUnet paper \cite{isensee2021nnu}. 

Three public datasets are used in this experiment. The Heart MRI dataset \cite{simpson2019large} has 20 labeled 3D images: 16 for training, 1 for validation and 3 for testing. The median shape of each 3D image is 115 $\times$ 320 $\times$ 320, of which 115 is the number of slices. In this experiment, only 2D segmentation is considered, so the input of the model is one slice. The batch size (40), input image size (320 $\times$ 256) are self-configured by nnUnet for this data set. The model is trained for 50 epochs, where each epoch contains 50 iterations. Other training settings are the same as those in \cite{isensee2021nnu}. For SAT, we tried three different noise levels (5, 15, 25). The Hippocampus MRI dataset  \cite{simpson2019large} has 260 labeled 3D images: 208 for training, 17 for validation and 35 for testing. The median shape of each 3D image is 36 $\times$ 50 $\times$ 35, where 36 is the number of slices. The batch size (366), the input image size (56 $\times$ 40) and network structure are self-configured by nnUnet for this data set. The model is trained for 100 epochs, where each epoch has 50 iterations. Other training settings are the same as those in \cite{isensee2021nnu}. For SAT, we tried four different noise levels (1, 5, 10, 15). The Prostate MRI dataset \cite{simpson2019large} has 32 labeled 3D images: 25 for training, 2 for validation and 5 for testing. The median shape of each 3D image is $20 \times 320 \times\ 319$, where $20$ is the number of slices. The batch size ($32$), patch size ($320 \times 320$) and network structure are self-configured by nnUnet for this dataset. The model is trained for 50 epochs, where each epoch has 50 iterations. Other training settings are the same as those in \cite{isensee2021nnu}. For SAT, we tried three different noise levels (10, 20, 40).

\subsection{Evaluation on landmark detection}

In this experiment, we apply AMAT and SAT to the Multi-Task U-Net \cite{yao2020miss}. This model detects the landmarks by regressing both Gaussian heatmap and offset maps of the landmarks simultaneously. The loss function for this model has three loss terms, and therefore $\xi$ is a vector of size 3, in which each value is set to $mean+2 \cdot std$ of the corresponding loss term values (Section \ref{2.2}). The original loss function contains a Binary Cross-Entropy (BCE) loss term. We find that, by replacing the BCE loss with Dice loss, the robustness of the Multi-Task U-Net becomes much better (Section 4). The  Multi-Task U-Net trained with the original loss is denoted as "STD(BCE)", and the one trained with the Dice loss is denoted as "STD(Dice)". The two methods "SAT" and "AMAT" are applied to the "STD(Dice)" models. The metric for this experiment is "Mean Radial Error (MRE)" defined in \cite{yao2020miss}, which measures the Euclidean distance between true and predicted landmarks. MRE is scaled to the unit of mm. The dataset is created for cephalometric landmark detection in IEEE ISBI 2015 Challenge \cite{wang2016benchmark}. The dataset contains 400 cephalometric radiographs, which are officially split into 3 sets (Train, Test1, Test2): 150 for Train, 150 for Test1 and 100 for Test2. Because the performance of the models reported in \cite{yao2020miss} is poor on Test2, we only use Test1. Test1 is further split into 2 sets: 50 for validation and 100 for testing. Each radiograph has 19 manually labeled landmarks of clinical anatomical significance by two expert doctors. The average annotations by two doctors is considered the ground truth. The images are resized to $200 \times 160$ in our experiment. The model is trained for 500 epochs. The batch size is 8. Other training settings are the same as those in \cite{yao2020miss}. For SAT, we tried three different noise levels (1, 3, 5). 

\subsection{Evaluation on object detection}

In this experiment, we apply AMAT and SAT to the YOLO v5s object detector \cite{glennjocher20226222936}. The loss function has three loss terms, and therefore $\xi$ is a vector of size 3, of which each value is set as $mean+2 \cdot std$ of the corresponding loss term values (Section 2.2). The metric used for evaluation is mAP 0.5:0.95. The metric used for tuning the hyperparameter is Average Intersection over Union (IOU) of all detected objects (Section \ref{3.8}). The dataset is named BCCD, which is a small-scale dataset for blood cell detection \cite{BCCDDataset}. The dataset has 364 2D images, which is split into 3 sets: 294 for training, 33 for validation and 37 for testing. The blood cells on each image are from three classes: Red Blood Cell, White Blood Cell, and Platelets. Each image is resized to $320 \times 320$ in our experiment. The model is trained for 300 epochs. The batch size is 32. Other training settings are the same as \cite{glennjocher20226222936}. For SAT, we tried three different noise levels (1, 10, 15).

\subsection{Selection of step size  $\Delta_\varepsilon$}
\label{3.8}

AMAT has the hyperparameter,  $\Delta_\varepsilon$, that needs to be tuned, and a grid search on the validation set can be conducted to decide the best value. The metric for hyperparameter selection is $AUC_{robust}$, borrowed from \cite{ma2022regularization}:
\begin{equation}
    AUC_{robust} = \sqrt{ACC_{clean} \times AUC_{||\tilde{x}-x||_p \leq \epsilon_{max}}}
\end{equation}

where $ACC_{clean}$ is the model accuracy on clean data (i.e., dice score for image segmentation, MRE for landmark detection, and IOU for object detection). $ AUC_{||\tilde{x}-x||_p \leq \epsilon_{max}}$ is the normalized area under the curve of accuracy vs. noise level, which is the average accuracy across a range of noise levels, i.e.$||\tilde{x}-x||_p \leq \epsilon_{max}$. Here, $\epsilon_{max}$ is the max noise level of the adversarial attack. $\tilde{x}$ and $x$ are defined in Algorithm \ref{algo1}. Therefore, $AUC_{robust}$ is the overall performance measure that takes both accuracy on clean data and robustness into consideration. The detailed grid search results are shown in the Appendix \ref{A1}. The configurations are the same as that mentioned in Section \ref{3}. Appendix \ref{A1} shows that the performance of AMAT is not significantly sensitive to the choice of $\Delta_\varepsilon$, except too small $\Delta_\varepsilon$. This is because AMAT with too small $\Delta_\varepsilon$ may need more training epochs to converge, while the grid research has a fixed number of epochs for each run.

From Table \ref{D2grid}, \ref{D4grid}, \ref{D5grid}, \ref{LDgrid}, \ref{ODgrid} in Appendix \ref{A1}, the optimal $\Delta_\varepsilon$ of AMAT for the experiments, Heart, Hippocampus, Prostate, Cephalometric and BCCD are 5, 0.5, 3, 1 and 3.5.

\section{Results and Discussions}

\subsection{Segmentation results on the three datasets (Heart, Hippocampus, and Prostate:}

\begin{table}[H]
\scriptsize
\centering
  \caption{ADI of nnUnet on Heart test set: "S. Dice" denotes the ADI on clean data; "P. $\epsilon$" denotes PGD attack with noise level $\epsilon$; "IF. $\epsilon$" denotes IFGSM attack with noise level $\epsilon$; "Avg." denotes average performance on noisy data}
  \label{D2}
  \begin{tabular}{l|l|llll|llll}
    \toprule
    Methods & S. Dice&P. 5&P. 10&P. 15& Avg.&IF. 5&IF. 10&IF. 15& Avg.\\
    \midrule
AMAT&	\textbf{0.9148}&	0.8476&	0.7594&	0.6588&	\textbf{0.7553}&	0.8834&	0.8532&	0.8255&	\textbf{0.8540}\\
									
PGD25&	0.6652&	0.6175&	0.5711&	0.5355&	0.5747&	0.6410&	0.6173&	0.6017&	0.6200\\
									
PGD15&	0.7188&	0.6579&	0.5869&	0.5194&	0.5881&	0.6933&	0.6640&	0.6329&	0.6634\\
									
PGD5&	0.7525&	0.6565&	0.5542&	0.4617&	0.5575&	0.7087&	0.6667&	0.6370&	0.6708\\
\midrule									
STD&	0.7954&	0.5197&	0.1997&	0.0172&	0.2455&	0.6930&	0.6238&	0.5779&	0.6316\\

    \bottomrule
  \end{tabular}
\end{table}

\begin{table}[H]
\scriptsize
\centering
  \caption{ADI of nnUnet on Hippocampus test set: "S. Dice" denotes the ADI on clean data; "P. $\epsilon$" denotes PGD attack with noise level $\epsilon$; "IF. $\epsilon$" denotes IFGSM attack with noise level $\epsilon$; "Avg." denotes average performance on noisy data}
  \label{D4}
  \begin{tabular}{l|l|llll|llll}
    \toprule
    Methods & S. Dice&P. 2&P. 6&P. 10& Avg.&IF. 2&IF. 6&IF. 10& Avg.\\
    \midrule
AMAT&	\textbf{0.7845}&	0.7525&	0.6752&	0.5580&	\textbf{0.6619}&	0.7587&	0.7280&	0.6968&	\textbf{0.7278}\\
									
PGD15&	0.7112&	0.6762&	0.5938&	0.4833&	0.5844&	0.6891&	0.6600&	0.6276&	0.6589\\
									
PGD10&	0.7175&	0.6768&	0.5953&	0.4707&	0.5809&	0.6949&	0.6599&	0.6232&	0.6593\\
									
PGD5&	0.7305&	0.6866&	0.5636&	0.4213&	0.5572&	0.7013&	0.6580&	0.6195&	0.6596\\
									
PGD1&	0.7510&	0.6616&	0.4192&	0.1752&	0.4187&	0.7067&	0.6303&	0.5693&	0.6354\\
\midrule									
STD&	0.7589&	0.5217&	0.0379&	0.0015&	0.1870&	0.6572&	0.5105&	0.3870&	0.5182\\

    \bottomrule
  \end{tabular}
\end{table}

\begin{table}[H]
\scriptsize
\centering
  \caption{ADI of nnUnet on Prostate test set: "S. Dice" denotes the ADI on clean data; "P. $\epsilon$" denotes PGD attack with noise level $\epsilon$; "IF. $\epsilon$" denotes IFGSM attack with noise level $\epsilon$; "Avg." denotes average performance on noisy data}
  \label{D5}
  \begin{tabular}{l|l|llll|llll}
    \toprule
    Methods & S. Dice&P. 10&P. 20&P. 40& Avg.&IF. 10&IF. 20&IF. 40& Avg.\\
    \midrule
AMAT&	\textbf{0.8336}&	0.7991&	0.7572&	0.6528&	\textbf{0.7364}&	0.8180&	0.8014&	0.7649&	\textbf{0.7948}\\
									
PGD40&	0.6152&	0.5423&	0.4614&	0.3085&	0.4374&	0.5817&	0.5453&	0.4768&	0.5346\\
									
PGD20&	0.6640&	0.5606&	0.4510&	0.2779&	0.4298&	0.6181&	0.5698&	0.4772&	0.5550\\
									
PGD10&	0.6569&	0.5405&	0.4034&	0.2063&	0.3834&	0.5981&	0.5522&	0.4676&	0.5393\\
	\midrule								
STD&	0.7418&	0.3012&	0.1428&	0.0455&	0.1632&	0.5252&	0.4138&	0.2796&	0.4062\\

    \bottomrule
  \end{tabular}
\end{table}

From Table \ref{tdvi}, the TVDI values of three "STD" models are similar to those in \cite{isensee2021nnu}, which indicates that the number of training epochs is enough for the models to converge. From Table \ref{D2}, Table \ref{D4} and Table \ref{D5}, we have the following observations. (1) AMAT has the best performance: the "AMAT" models have the highest average ADI on noisy data; the "AMAT" model has the highest ADI on clean data, even higher than the "STD" model. (2) Under the 100-PGD attack, the ADI of the "STD" models drops to almost 0 on the noise level 15 on the Heart dataset, on the noise level 10 on the Hippocampus dataset, and on the noise level 40 on the Prostate dataset. (3) The "$SAT\varepsilon$" models with smaller $\varepsilon$ have better performance on lower-noise data, but worse performance on higher-noise data. (4) The "$SAT\varepsilon$" models with larger $\varepsilon$ have better performance on higher-noise data, but worse performance on lower-noise data and clean data.

\begin{center}
\scriptsize
\begin{minipage}[ht]{0.22\textwidth}
\centering
\captionof{table}{TDVI of "STD" models on three datasets}
\label{tdvi}
\begin{tabular}{l|l}
\hline
Task & TVDI\\
\hline
Heart &0.92  \\
Hippocampus &0.88\\
Prostate &0.82\\
\hline
\end{tabular}
\end{minipage}
\hfill
\begin{minipage}[ht]{0.7\textwidth}

\centering
\captionof{table}{Time cost for each training epoch (Seconds): Columns 1-3 are for segmentation; Column 4 is for landmark detection; Column 5 is for blood cell detection.}
\label{segtime}
\begin{tabular}{l|lllll}
\hline
- &Heart & Hippocampus & Prostate & Landmark & BCCD\\
\hline
STD &31 &10 &20& 12& 6\\
SAT &476 &114 &465& 50 & 30\\
AMAT &480 &120 &467& 53& 35\\
\hline
\end{tabular}
\end{minipage}
\end{center}

\subsection{Landmark detection results on the Cephalometric dataset}

\begin{table}[H]
\scriptsize
\centering
  \caption{MRE (mm) of Multi-Task Unet on Cephalometric test set: "MRE" denotes the MRE on clean data; "P. $\epsilon$" denotes PGD attack with noise level $\epsilon$; "IF. $\epsilon$" denotes IFGSM attack with noise level $\epsilon$; "Avg." denotes average performance on noisy data. MRE is the lower the better.}
  \label{LD}
  \begin{tabular}{l|l|llll|llll}
    \toprule
    Methods & MRE&P. 0.5&P. 1&P. 2& Avg.&IF. 0.5&IF. 1&IF. 2& Avg.\\
    \midrule
AMAT&	\textbf{1.773}&	1.891&	2.044&	2.415&	\textbf{2.117}&	1.892&	2.037&	2.391&	\textbf{2.107}\\    

SAT1&	1.814&	1.948&	2.057&	2.457&	2.154&	1.948&	2.060&	2.427&	2.145\\
									
SAT3&	2.108&	2.187&	2.270&	2.477&	2.311&	2.187&	2.274&	2.469&	2.310\\
									
SAT5&	2.177&	2.263&	2.342&	2.543&	2.383&	2.264&	2.343&	2.545&	2.384\\
								
	\midrule								
STD(BCE)&	1.545&	4.345&	14.295&	38.868&	19.169&	3.929&	10.910&	35.015&	16.618\\
									
STD(Dice)&	1.623&	3.378&	8.289&	32.820&	14.829&	3.077&	6.245&	21.285&	10.202\\

    \bottomrule
  \end{tabular}
\end{table}

From  Table \ref{LD}, observations are as follows. (1) AMAT has the best performance: The "AMAT" model has the smallest Avg. MRE on noisy data; The "AMAT" model has the smallest Avg. MRE on clean data among all defense methods. (2) The "STD(Dice)" model is better than the "STD(BCE)" model. But both of them are very vulnerable to 100-PGD and 10-IFGSM adversarial attacks. (3) The "$SAT\varepsilon$" models with smaller $\varepsilon$ have better performance on lower-noise data and clean data, but not as good as AMAT. (4) The "$SAT\varepsilon$" models with larger $\varepsilon$ have worse performance on lower-noise data and clean data.

\subsection{Blood cell detection results on the BCCD dataset} 

\iffalse
\begin{table}[H]
\scriptsize
\centering
  \caption{Avg. IOU of YOLO V5 on BCCD test set: "IOU" denotes the average IOU on clean data; "P. $\epsilon$" denotes PGD with noise level $\epsilon$; "IF. $\epsilon$" denotes IFGSM attack with noise level $\epsilon$; "Avg." denotes average performance on noisy data}
  \label{OD1}
  \begin{tabular}{l|l|llll|llll}
    \toprule
    Methods & IOU &P. 5&P. 10&P. 15& Avg.&IF. 5&IF. 10&IF. 15& Avg.\\
    \midrule
AMAT&	\textbf{0.7796}&	0.5799&	0.3647&	0.2832&\textbf{0.4093}&	0.5839&	0.3876&	0.3013 &\textbf{0.4243}\\

SAT1&	0.7407&	0.4417&	0.3413&	0.2773&	0.3534&	0.4534&	0.3509&	0.3129&	0.3724\\
									
%SAT10&	0.7039&	0.5835&	0.4064&	0.2727&	0.4208&	0.5831&	0.4139&	0.2847&	0.4273\\
									
SAT15&	0.5546&	0.4595&	0.3679&	0.2943&	0.3739&	0.4600&	0.3688&	0.2953&	0.3747\\
\midrule
STD&	0.7602&	0.1560&	0.1079&	0.0841&	0.1160&	0.1825&	0.1184&	0.1004&	0.1338\\							
    \bottomrule
  \end{tabular}
\end{table}
\fi

\begin{table}[H]
\scriptsize
\centering
  \caption{mAP0.5:0.95 of YOLO V5 on BCCD test set: "mAP" denotes the mAP0.5:0.95 on clean data; "P. $\epsilon$" denotes PGD with noise level $\epsilon$; "IF. $\epsilon$" denotes IFGSM attack with noise level $\epsilon$; "Avg." denotes average performance on noisy data}
  \label{OD2}
  \begin{tabular}{l|l|llll|llll}
    \toprule
    Methods & mAP &P. 5&P. 10&P. 15& Avg.&IF. 5&IF. 10&IF. 15& Avg.\\
    \midrule

AMAT&	\textbf{0.5989}&	0.3341&	0.2017&	0.1428&	\textbf{0.2262}&	0.3350&	0.2106&	0.1511&	\textbf{0.2322}	\\								
SAT1&	0.5810&	0.2845&	0.1467&	0.0815&	0.1709&	0.2954&	0.1762&	0.1228&	0.1981\\
									
SAT10&	0.4722&	0.2988&	0.1869&	0.0765&	0.1874&	0.2988&	0.1868&	0.0808&	0.1888\\
									
SAT15&	0.2565&	0.2013&	0.1375&	0.0572&	0.1320&	0.2015&	0.1354&	0.0603&	0.1324\\
\midrule
STD&	0.5993&	0.0072&	0.0007&	0.0005&	0.0028&	0.0456&	0.0102&	0.0082&	0.0213	\\

    \bottomrule
  \end{tabular}
\end{table}

From Table \ref{OD2}, observations are as follows. (1) AMAT has the best performance: The "AMAT" model has the best average performance on noisy data; The "AMAT" model has the best performance on clean data among all defense methods, very close to "STD". (2) The performance of the "STD"  model drops to 0 on very small noise levels. (3) "$SAT\varepsilon$" models with smaller $\varepsilon$ has better performance on lower-noise data and clean data, but not as good as AMAT. (4)"$SAT\varepsilon$" models with larger $\varepsilon$ has worse performance on lower-noise data and clean data.

\subsection{Discussion}

We have those observations from the experiments. (1) Our AMAT method has the best performance: AMAT has the best average performance on noisy data, compared to STD and SAT; AMAT has better performance on clean data, compare to SAT. In the image segmentation of Heart, Hippocampus and Prostate, AMAT even outperforms STD on clean data. This suggests that properly crafted adversarial training noise may not have to harm the model's performance. In other words, the trade-off between accuracy and robustness can be lifted for this task: we do not need to sacrifice accuracy (on clean data) in exchange for robustness. This is also corroborated by a similar observation in a whole-image classification task \cite{zhang2021geometry}. (2) Table \ref{segtime} shows that AMAT has almost the same training time cost as SAT. (3) The noises measured by L2 norm in the experiments are very small on each pixel, but the performance of the "STD" models drops to almost zero on small noise levels, which indicates the 100-PGD is a very strong and covert threat to segmentation and detection tasks. (4) The "$SAT\varepsilon$" models with larger $\varepsilon$ have worse performance on lower-noise data, and the "$SAT\varepsilon$" models with smaller $\varepsilon$ have poor performance on higher-noise data. This indicates that a fixed and unified level of noise is inadequate for robust DNN training. For life-critical medical image analysis applications, the DNN models should have a good performance on both clean data and noisy data. Thus, "AMAT" with adaptive training noises is more suitable than "SAT$\varepsilon$" with a fixed and unified noise level $\varepsilon$. 

\section{Conclusion}

We propose an adaptive adversarial training method, AMAT, to make DNNs robust against adversarial attacks/noises in medical image segmentation, landmark detection, and object detection tasks, and the key idea is to adaptively adjust the training noises for individual training samples, so that the trained DNN model will have good resistance to the adversarial noises while maintaining good accuracy on clean data. We apply our proposed method to three state-of-the-art DNNs on five publicly available medical image datasets. The experiment results show that our AMAT method outperforms the SAT method in both adversarial robustness and prediction accuracy on clean data, while having almost the same training time cost as SAT. We hope our approach may facilitate the development of robust DNNs for more medical applications in the future.

%\acks{Acknowledgements should go at the end, before appendices and references. You can uncomment this for the camera-ready version on paper acceptance.}

%\bibliographystyle{plain}
\bibliography{mybib}

\appendix

\section{Selection of $\Delta_\varepsilon$}
\label{A1}

\begin{table}[H]
\scriptsize
\centering
  \caption{ADI of nnUnet on Heart validation set}
  \label{D2grid}
  \begin{tabular}{l|llll|l}
    \toprule
    Noise level & 0 & 5 & 10 & 15 & $AUC_{Robust}$\\
    \midrule
$\Delta_\varepsilon$ = 1&	0.8902&	0.8373&	0.7702&	0.6930& 0.8438\\

$\Delta_\varepsilon$ = 3&	0.9067&	0.8520&	0.7965&	0.7363& 0.8640\\

$\Delta_\varepsilon$ = 5&	0.9101&	0.8673&	0.8205&	0.7681& \textbf{0.8755}\\

$\Delta_\varepsilon$ = 7&	0.9076&	0.8657&	0.8193&	0.7757& 0.8743\\

$\Delta_\varepsilon$ = 9&	0.9037&	0.8608&	0.8181&	0.7675& 0.8703\\

    \bottomrule
  \end{tabular}
\end{table}

\begin{table}[H]
\scriptsize
\centering
  \caption{ADI of nnUnet on Hippocampus validation set}
  \label{D4grid}
  \begin{tabular}{l|llll|l}
    \toprule
    Noise level & 0 & 2 & 6 & 10 & $AUC_{Robust}$\\
    \midrule
$\Delta_\varepsilon$ = 0.1&	0.7860&	0.7460&	0.6482&	0.5004&	0.7212\\
					
$\Delta_\varepsilon$ = 0.3&	0.7733&	0.7418&	0.6655&	0.5562&	0.7237\\
					
$\Delta_\varepsilon$ = 0.5&	0.7753&	0.7407&	0.6709&	0.5590&	\textbf{0.7260}\\
					
$\Delta_\varepsilon$ = 0.7&	0.7718&	0.7408&	0.6692&	0.5522&	0.7231\\
					
$\Delta_\varepsilon$ = 0.9&	0.7665&	0.7374&	0.6650&	0.5640&	0.7202\\
					
$\Delta_\varepsilon$ = 1.1&	0.7685&	0.7373&	0.6681&	0.5583&	0.7213\\
					
$\Delta_\varepsilon$ = 1.3&	0.7668&	0.7371&	0.6678&	0.5663&	0.7211\\
					
$\Delta_\varepsilon$ = 1.5&	0.7679&	0.7364&	0.6643&	0.5640&	0.7206\\
    \bottomrule
  \end{tabular}
\end{table}

\begin{table}[H]
\scriptsize
\centering
  \caption{ADI of nnUnet on Prostate validation set}
  \label{D5grid}
  \begin{tabular}{l|llll|l}
    \toprule
    Noise level & 0 & 10 & 20 & 40 & $AUC_{Robust}$\\
    \midrule
$\Delta_\varepsilon$ = 1&	0.8351&	0.7981&	0.7520&	0.6339&	0.7884\\
					
$\Delta_\varepsilon$ = 3&	0.8271&	0.7972&	0.7572&	0.6637&	\textbf{0.7890}\\
					
$\Delta_\varepsilon$ = 5&	0.8018&	0.7756&	0.7446&	0.6757&	0.7715\\
					
$\Delta_\varepsilon$ = 7&	0.8119&	0.7872&	0.7550&	0.6774&	0.7807\\
					
$\Delta_\varepsilon$ = 9&	0.8096&	0.7835&	0.7487&	0.6731&	0.7772\\
					
$\Delta_\varepsilon$ = 11&	0.8033&	0.7781&	0.7481&	0.6762&	0.7734\\
    \bottomrule
  \end{tabular}
\end{table}

\begin{table}[H]
\scriptsize
\centering
  \caption{MRE (mm) of Multi-Task Unet on Cephalometric validation set}
  \label{LDgrid}
  \begin{tabular}{l|llll|l}
    \toprule
    Noise level & 0 & 0.5 & 1 & 2 & $AUC_{Robust}$\\
    \midrule
$\Delta_\varepsilon$ =0.1&	1.637&	1.906&	2.247&	3.870&	0.5218\\
					
$\Delta_\varepsilon$ =0.5&	1.735&	1.918&	2.103&	2.817&	0.5036\\
					
$\Delta_\varepsilon$ =1.0&	1.773&	1.892&	2.044&	2.426&	\textbf{0.4947}\\
					
$\Delta_\varepsilon$ =1.5&	1.893&	1.992&	2.117&	2.388&	0.5183\\
					
$\Delta_\varepsilon$ =2.0&	1.978&	2.081&	2.195&	2.420&	0.5385\\

    \bottomrule
  \end{tabular}
\end{table}

\begin{table}[H]
\scriptsize
\centering
  \caption{Avg. IOU of YOLO V5 on BCCD validation set}
  \label{ODgrid}
  \begin{tabular}{l|llll|l}
    \toprule
    Noise level & 0 & 5 & 10 & 15 & $AUC_{Robust}$\\
    \midrule
$\Delta_\varepsilon$ =0.5&	0.7672&	0.5444&	0.3509&	0.2640&	0.6007\\
					
$\Delta_\varepsilon$ =1.0&	0.7633&	0.5539&	0.3570&	0.2718&	0.6029\\
					
$\Delta_\varepsilon$ =1.5&	0.7650&	0.5657&	0.3599&	0.2697&	0.6066\\
					
$\Delta_\varepsilon$ =2.0&	0.7636&	0.5683&	0.3546&	0.2718&	0.6055\\
					
$\Delta_\varepsilon$ =2.5&	0.7777&	0.5720&	0.3699&	0.2601&	0.6154\\
					
$\Delta_\varepsilon$ =3.0&	0.7767&	0.5775&	0.3577&	0.2686&	0.6144\\
					
$\Delta_\varepsilon$ =3.5&	0.7796&	0.5776&	0.3655&	0.2792&	\textbf{0.6186}\\
					
$\Delta_\varepsilon$ =4.0&	0.7779&	0.5733&	0.3460&	0.2632&	0.6110\\
					
$\Delta_\varepsilon$ =4.5&	0.7745&	0.5501&	0.3349&	0.2573&	0.6014\\
					
$\Delta_\varepsilon$ =5.0&	0.7717&	0.5572&	0.3329&	0.2319&	0.5984\\

    \bottomrule
  \end{tabular}
\end{table}

\section{Formulars}
\label{A2}
\begin{equation}
TVDI = \frac{2\times \sum_i^n{TP_i}}{2\times \sum_i^n{TP_i} + \sum_i^n{FP_i} +\sum_i^n{FN_i} }
\end{equation} 
The notations are the same as equation (2) in Section 3.1.

\section{Image samples}

\begin{center}
\begin{minipage}[ht]{\textwidth}
\centering
\includegraphics[width=0.7\linewidth]{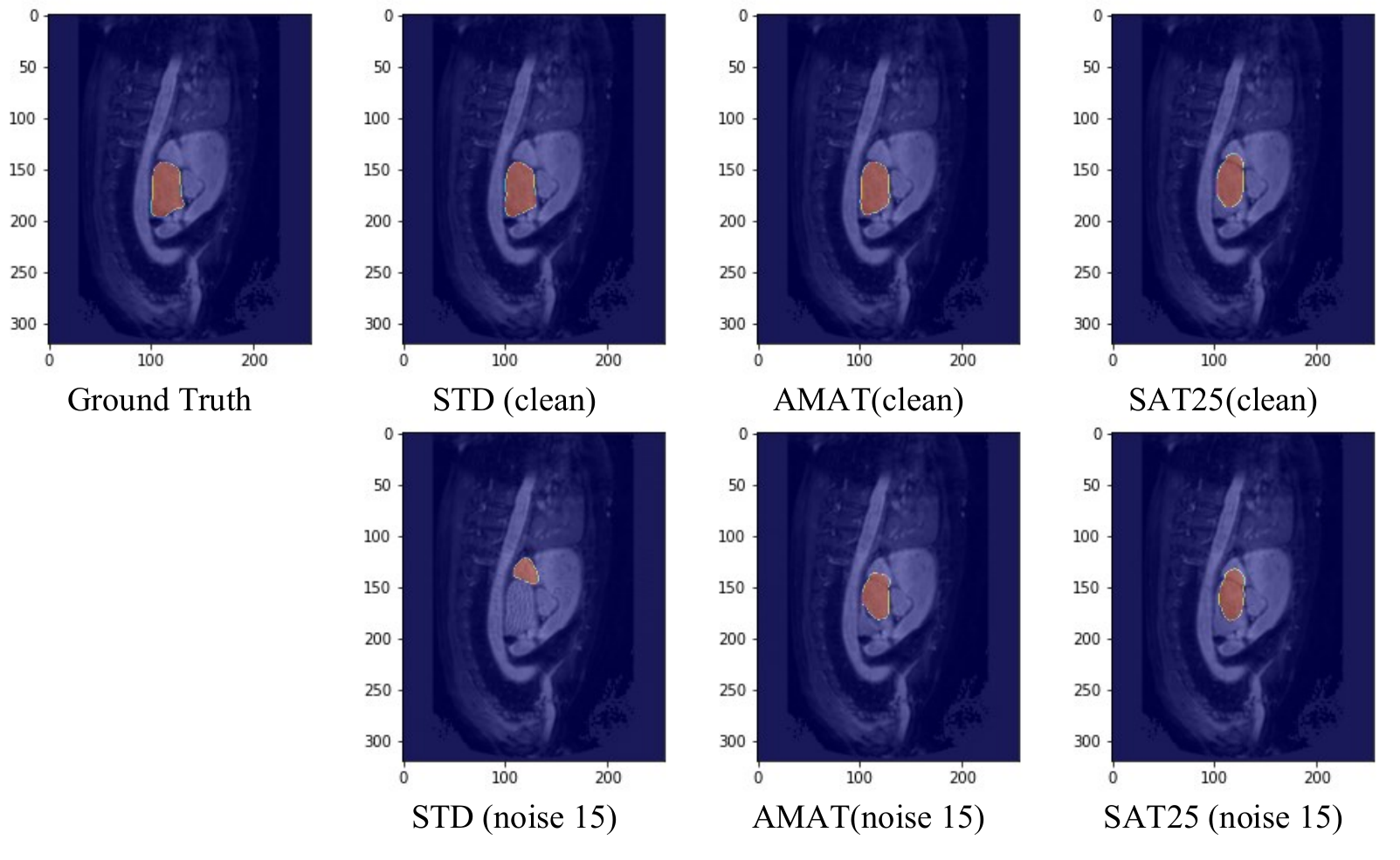}
    \captionof{figure}{Examples of adversarial attacks against Heart MRI dataset (L2 norm)}
    \label{intro}
 \end{minipage}
 \end{center}
 
  \begin{center}
\begin{minipage}[ht]{\textwidth}
\centering
\includegraphics[width=0.7\linewidth]{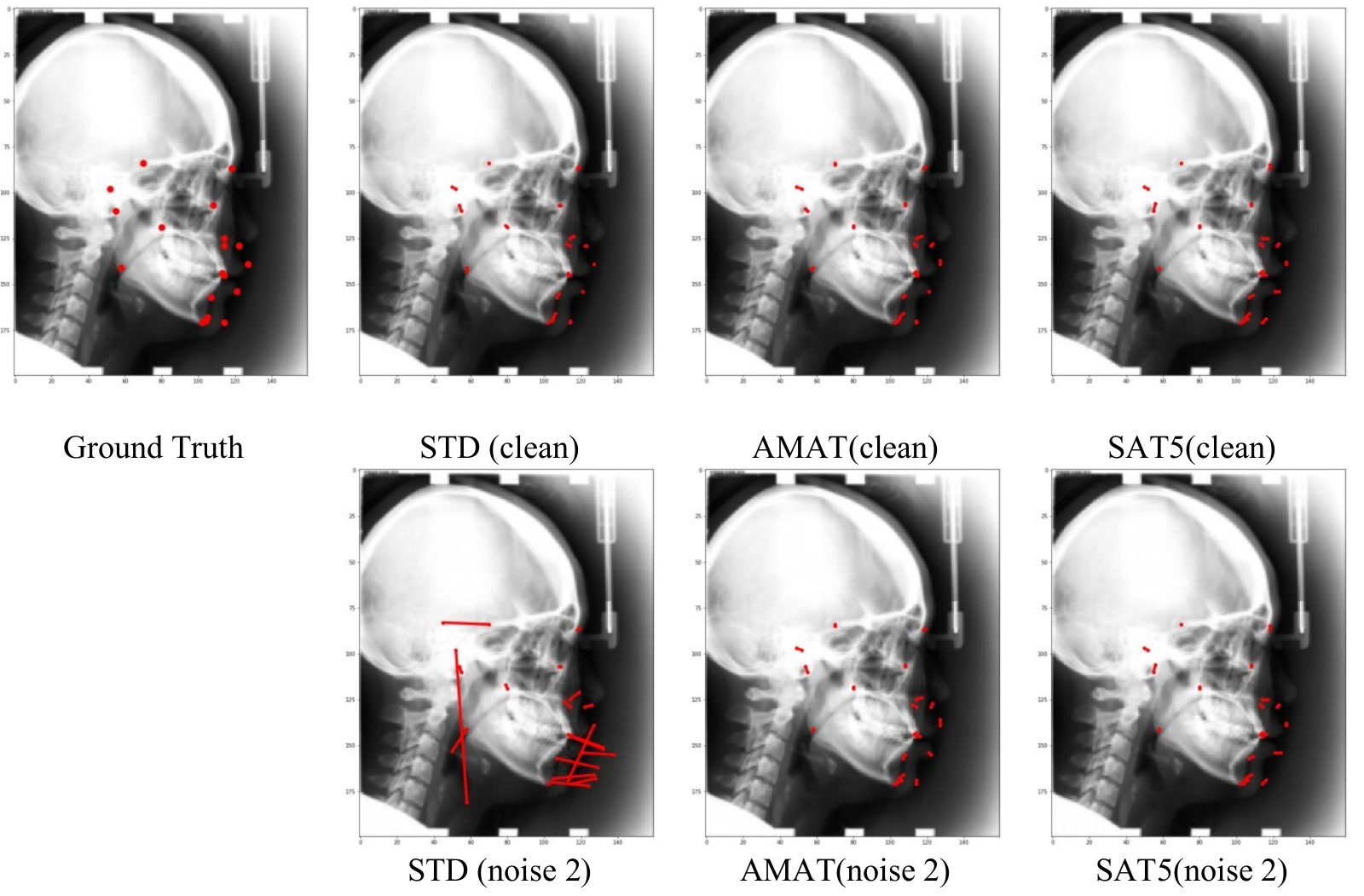}
    \captionof{figure}{Examples of adversarial attacks against cephalometric landmark detection (L2 norm). The red lines indicates the distance between ground truths and predicted locations}
    \label{intro}
 \end{minipage}
 \end{center}
 
   \begin{center}
\begin{minipage}[ht]{\textwidth}
\centering
\includegraphics[width=0.98\linewidth]{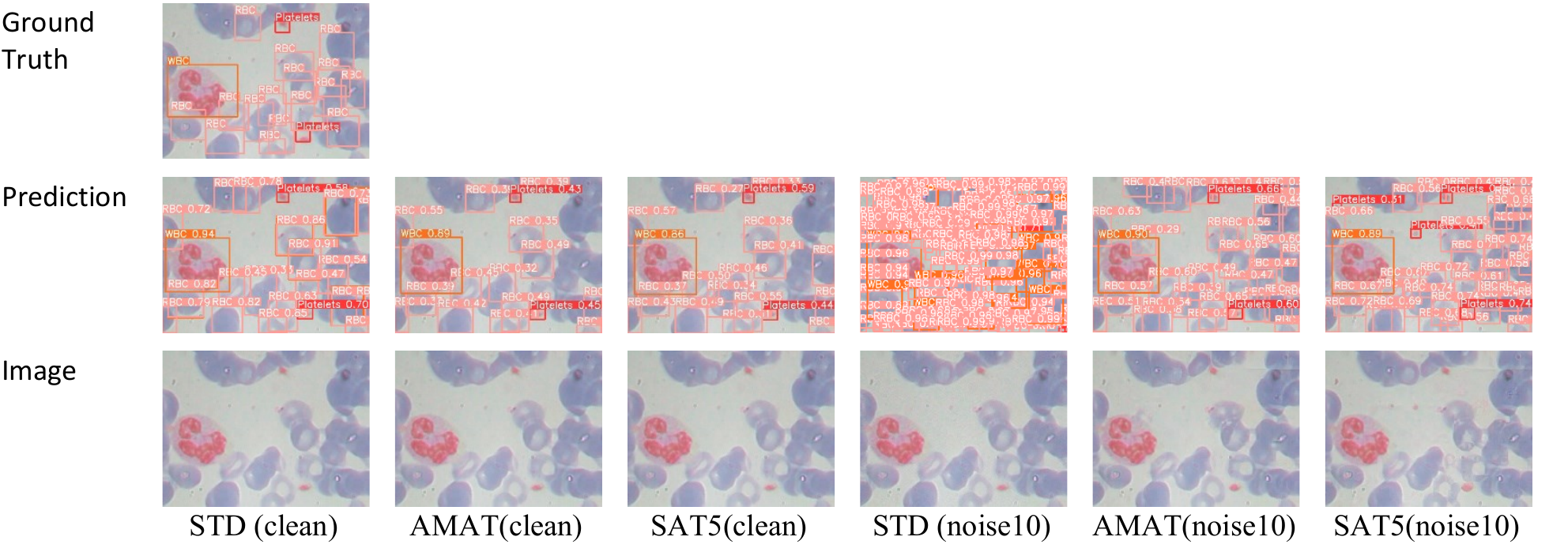}
    \captionof{figure}{Examples of adversarial attacks against blood cell detection  (L2 norm)}
    \label{intro}
 \end{minipage}
 \end{center}

\end{document}